\documentclass{egpubl}
\CGFccby

\usepackage[T1]{fontenc}
\usepackage{dfadobe}

\usepackage[T1]{fontenc}
\usepackage{times} 

\usepackage{cite}  
\BibtexOrBiblatex
\electronicVersion
\PrintedOrElectronic
\ifpdf \usepackage[pdftex]{graphicx} \pdfcompresslevel=9
\else \usepackage[dvips]{graphicx} \fi

\usepackage{egweblnk} 

\usepackage{amsmath,amssymb}
\usepackage{algorithm}
\usepackage{algorithmic}

\begin{document}

\title {PhysMorph-GS: Render-Guided Volumetric Morphing with Differentiable Physics}

\author[C. Y. Song \& D. Hyde]
{\parbox{\textwidth}{\centering C.\,Y. Song\thanks{Corresponding Author}$^{1}$\orcid{0000-0002-2495-1383}
        and D. Hyde$^{1}$\orcid{0009-0004-4950-5533} 
        }
        \\
{\parbox{\textwidth}{\centering $^1$Vanderbilt University, Nashville TN, USA
       }
}
}

\teaser{
 \includegraphics[width=0.85\linewidth]{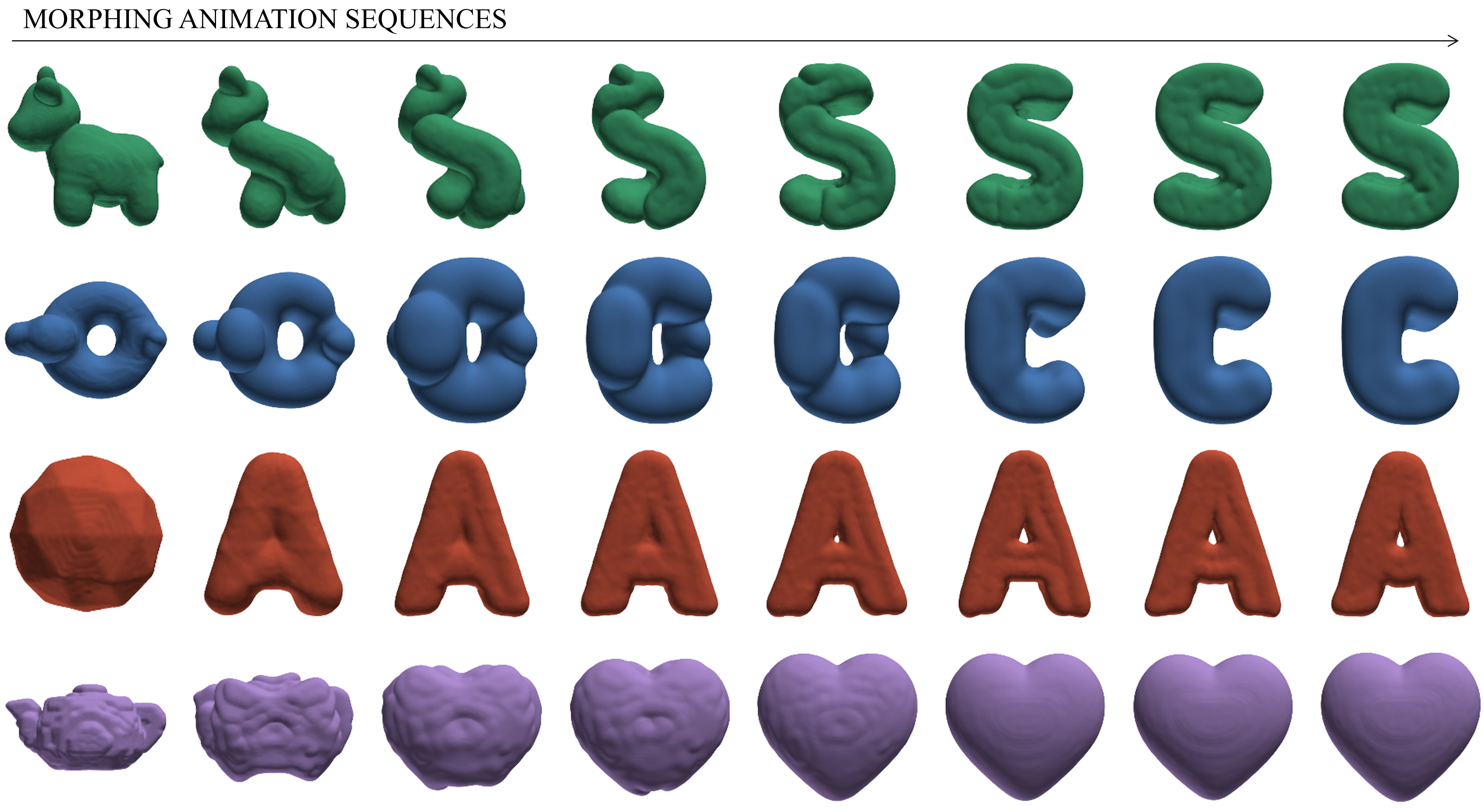}
 \centering
  \caption{Physically plausible morphing guided by differentiable 3D Gaussian splatting. Each MPM particle simultaneously serves as a Gaussian primitive via $\boldsymbol{\Sigma} = \mathbf{F}\boldsymbol{\Sigma}_0\mathbf{F}^\top$. Rendering supervision acts exclusively on the deformation gradient $\mathbf{F}$ while positions are governed by physics. Chamfer-guided plasticity migrates the elastic rest state toward the target, converting restoring forces from adversary to ally.}
 \label{fig:teaser}
}

\maketitle
\begin{abstract}
Differentiable particle-based simulation can produce physically plausible motion, but target-driven volumetric shape morphing remains underconstrained: physics-only mass matching captures coarse global structure yet struggles with fine geometric detail, while naive image-space coupling destabilizes elastic dynamics.
We present PhysMorph-GS, a render-guided morphing framework that couples material point method simulation with differentiable 3D Gaussian splatting.
The key idea is to inject visual supervision through the deformation gradient $\mathbf{F}$ rather than particle positions, so render gradients act as control-space guidance while trajectories remain governed by physics.
We further introduce phased Chamfer-guided plasticity that delays rest-state migration until coarse structure has formed; in practice, rendering is evaluated on a surface-focused particle subset for efficiency and gradient concentration.
Relative to a physics-only baseline, our method reduces silhouette error by 25.8\%, 10.8\%, and 49.9\% on representative examples, with the largest gains on models with thin features.
These results suggest that the main challenge in render-guided differentiable morphing is not simply adding stronger image losses, but injecting visual guidance in a way that remains compatible with elastic simulation.
We further observe that plasticity-driven rest-state migration drives different sources toward a shared target-determined attractor, distinguishing physics-based morphing from interpolation between registered shape pairs.


\end{abstract}

\section{Introduction}
\label{sec:intro}

Shape morphing, the process of smoothly deforming one 3D shape into another, is a fundamental operation in computer graphics with applications in visual effects, product design exploration through shape variations, and simulation-based content generation. The ideal morphing system would produce transitions that are both physically plausible (respecting material properties and conservation laws) and visually faithful to a desired target shape.

Prior work on this problem has evolved along two main directions. First, geometric approaches have laid the foundation for shape morphing by interpolating mesh vertex positions~\cite{alexa2023rigid} or blending implicit representations~\cite{turk2005shape}, achieving smooth visual transitions. However, since these methods do not model material mechanics, intermediate shapes may self-intersect, violate volume conservation, or exhibit physically implausible folding. Second, physics-based deformable models~\cite{terzopoulos1988deformable,bouaziz2023projective,muller2004point} have made significant contributions by simulating material response to restore physical plausibility, but they provide no mechanism to actively guide the deformation toward a specific target shape, as the final configuration is determined by the material model and boundary conditions alone.

Recent advances in differentiable physics bridge this gap by enabling gradient-based optimization of simulation parameters to match desired outcomes~\cite{hu2019difftaichi,8794333,du2021diffpd}. In the context of shape morphing, Xu et al.~\cite{11088224} demonstrated that a differentiable material point method (MPM) can morph one volumetric shape into another by optimizing a control deformation field that steers the simulation toward a target mass distribution. The result is a physically plausible animation that approximately matches the target shape.

However, the physics-only formulation has two fundamental limitations. First, it optimizes a volumetric mass-matching loss on the Eulerian grid, providing only coarse shape guidance; fine geometric details such as thin protrusions, concavities, and surface features are difficult to capture with grid-based losses alone. Second, particle-based simulations produce point cloud outputs, requiring surface reconstruction or conversion to a renderable representation for visual quality assessment. These two limitations, the absence of fine-grained shape guidance and the need for a renderable representation, naturally motivate simultaneous rendering and physics supervision. In this work, we integrate 3D Gaussian splatting (3DGS)~\cite{kerbl20233d} to compare multi-view silhouettes and depth against target images, obtaining dense pixel-level shape feedback that supplements the volumetric physics loss. 3DGS is particularly well-suited because each particle maps directly to a Gaussian primitive without explicit surface reconstruction, enabling differentiable rendering. Physics-integrated variants such as PhysGaussian~\cite{xie2024physgaussian} have demonstrated that this coupling is effective in practice, using the particle deformation gradient to warp Gaussian covariance.

While combining rendering supervision with physics simulation is straightforward in principle, the results depend critically on how and when the rendering signal is injected into the physics loop. Our investigation reveals two failure modes of naive coupling. First, injecting rendering gradients into particle positions requires ad-hoc cross-space scaling (the position gradient lives in configuration space while the physics update is driven by stresses) and is empirically unstable at gains that $\mathbf{F}$-space injection handles without issue (Section~\ref{sec:ablation}). Second, coupling the target too aggressively in early frames, before the physics has established meaningful structural seeds, can drive the system toward incorrect basins, preventing the desired structural bifurcation from emerging (e.g., legs separating from a body).

We address both challenges through a render-guided framework with two core algorithmic components:

\begin{enumerate}
\item \textbf{Control-space guidance via $\mathbf{F}$.} Instead of injecting render gradients into particle positions, we route them exclusively through the deformation gradient $\mathbf{F}$ via a differentiable $\mathbf{F} \to \boldsymbol{\Sigma}$ mapping~\cite{irving2004invertible}. This converts multi-view alpha/depth supervision into smoothed control-space guidance: the render signal modifies the control deformation field, which indirectly influences particle trajectories through the physics solver's own forward simulation, respecting material properties and stability constraints rather than directly opposing elastic restoring forces.

\item \textbf{Chamfer-guided plasticity with phased coupling.}
Render injection provides lightweight $\mathbf{F}$-space guidance from the first frame, while Chamfer-guided plasticity activates only after a warmup of $k_0$ frames once coarse structural seeds have formed. Under the multiplicative decomposition $\mathbf{F} = \mathbf{F}_e \mathbf{F}_p$, updating $\mathbf{F}_p$ via Chamfer nearest-neighbor displacements migrates the elastic rest configuration toward the target, converting elastic restoring forces from an adversary into an ally that actively drives particles toward the target shape (Section~\ref{sec:chamfer}).

\end{enumerate}

For efficiency, rendering is evaluated on a surface-focused subset of particles identified via density reconstruction and augmented with a Chamfer-based proximity criterion; in our experiments, this detail can accelerate algorithm performance by upwards of 60\% (Section~\ref{sec:surface}). 

\begin{figure*}[t]
  \centering
  \includegraphics[width=0.9\linewidth]{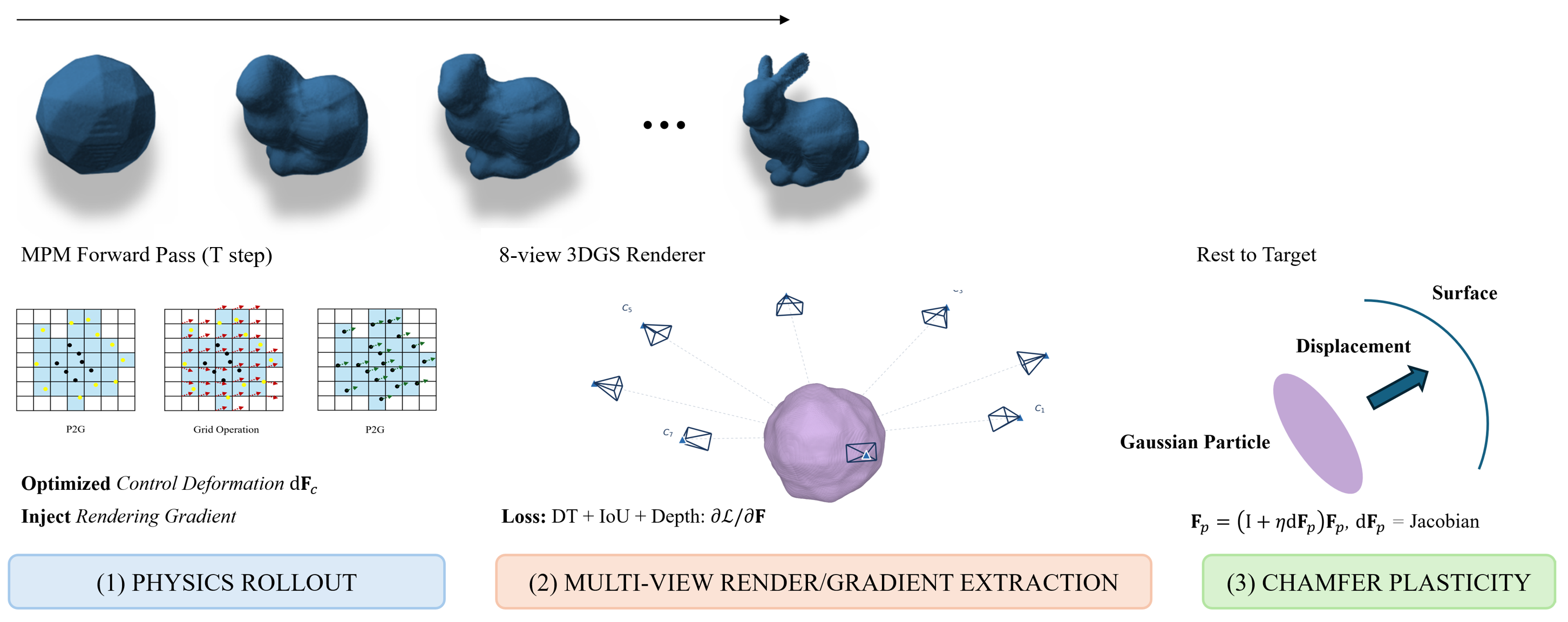}
  \caption{Pipeline overview. Each frame: (1)~physics rollout with F-only render penalty from the previous frame, (2)~multi-view rendering through the differentiable $\mathbf{F}\!\to\!\boldsymbol{\Sigma}$ chain, extracting $\partial \mathcal{L}/\partial\mathbf{F}$ for the next frame, and (3)~Chamfer-guided plasticity updating $\mathbf{F}_p$ to migrate the rest configuration. Positions are governed solely by physics; rendering acts only on $\mathbf{F}$ (covariance).}
  \label{fig:pipeline}
\end{figure*}

Our contributions are:
\begin{itemize}
\item \textbf{Control-space render guidance via MPM--Gaussian duality.} We establish a one-to-one correspondence between MPM particles and 3D Gaussian primitives via $\boldsymbol{\Sigma}_i = \mathbf{F}_i \boldsymbol{\Sigma}_0 \mathbf{F}_i^\top$, and inject multi-view render gradients exclusively through $\partial \mathcal{L}/\partial\mathbf{F}$ while zeroing $\partial \mathcal{L}/\partial\mathbf{x}$. Since $\partial \mathcal{L}/\partial\mathbf{F}$ is dimensionally aligned with the Piola–Kirchhoff stress that the particle-to-grid (P2G) step already integrates, the render signal composes directly with the physics update without cross-space magnitude matching, yielding stable render-guided optimization.

\item \textbf{Phased Chamfer-guided plasticity.} After a warmup period, we update the plastic deformation $\mathbf{F}_p$ using Chamfer nearest-neighbor displacements, migrating the elastic rest configuration toward the target. Combined with velocity damping, this turns achieved deformation into persistent progress instead of rebound, and leads to morphing trajectories that converge to a target-dominated attractor largely independent of the initial source shape (Section~\ref{sec:source_invariance}).

\item \textbf{Empirical analysis of render--physics coupling.} We provide ablations across velocity damping, render guidance, and target complexity, and show that the proposed coupling consistently improves over a physics-only baseline, with the largest gains on targets with fine structures.
\end{itemize}

For equivalent loss values obtained using a physics-only baseline, our full method reduces silhouette error by 25.8\%, 10.8\%, and 49.9\% on Bunny, Cow, and Duck, respectively.

Throughout the paper, we use \emph{physically plausible} in the graphics sense: trajectories are produced by an MPM simulator with a fixed constitutive model and stability constraints, rather than by unconstrained geometric interpolation. We do not claim exact correspondence to captured real-world material behavior.

\section{Related Work}
\label{sec:related}

\subsection{Shape Morphing}
Shape morphing has evolved along geometric, metric-based, and physics-based directions, each with inherent limitations. Geometric methods such as vertex interpolation~\cite{alexa2023rigid}, implicit blending~\cite{turk2005shape}, as-rigid-as-possible interpolation~\cite{alexa2023rigid}, and latent-space learning~\cite{groueix20183d,li2024deformnet} produce visually smooth transitions but lack material mechanics, leading to self-intersection and volume non-conservation. Metric-based optimization via Chamfer distance~\cite{fan2017point,song2026structuralfailurechamferdistance} or Wasserstein distance~\cite{peyre2019computational} can directly reduce shape dissimilarity, and Zhang et al.~\cite{zhang2022wassersplines} extended this to animation through neural vector fields along Wasserstein geodesics, yet neither guarantees physical plausibility of intermediate shapes. More broadly, these geometric and metric-based approaches either require dense correspondence between registered shape pairs (as in ARAP-style interpolation~\cite{alexa2023rigid}) or operate on abstract shape distributions without tracking material state, and none expose material state (e.g., elastic energy or plastic deformation) during the morphing trajectory itself. Physics-based deformable models~\cite{terzopoulos1988deformable,bouaziz2023projective,muller2004point} restore plausibility through material response but cannot actively guide deformation toward a specific target. Target-driven animation has been explored for fluids~\cite{manteaux2016space,10.1145/882262.882337} via keyframe control, but volumetric solid morphing with visual supervision remains underexplored. Xu et al.~\cite{11088224} demonstrated differentiable MPM-based morphing by optimizing a control deformation field, and we adopt this as our physics-only baseline.

\subsection{Material Point Method}
MPM~\cite{sulsky1994particle} discretizes continua as Lagrangian particles exchanging information through an Eulerian grid. It has been applied to diverse materials in graphics~\cite{stomakhin2013material,klar2016drucker,jiang2017anisotropic,wolper2019cd,10.1145/3414685.3417845,conservative_surface_tension}, and the APIC~\cite{10.1145/2766996} and MLS-MPM~\cite{hu2018moving} transfer schemes improved accuracy. The multiplicative decomposition $\mathbf{F} = \mathbf{F}_e \mathbf{F}_p$~\cite{lee1969elastic} separates elastic and plastic deformation for independent control; invertible formulations~\cite{2012-FixedCoratedElasticty} ensure robustness. Differentiable MPM (ChainQueen~\cite{8794333}, DiffTaichi~\cite{hu2019difftaichi}, DiffPD~\cite{du2021diffpd}) enables gradient-based optimization through simulation, and we build on Xu et al.~\cite{11088224}, who demonstrated differentiable MPM-based morphing via control deformation fields.

\subsection{Differentiable Rendering and Physics Integration}
Differentiable rendering~\cite{kato2020differentiable} began with point-based splatting~\cite{yifan2019differentiable}, Soft Rasterizer~\cite{liu2019softras}, and modular primitives~\cite{laine2020modular}, then expanded to neural implicit representations such as NeRF~\cite{mildenhall2020nerf,mueller2022instant} and NeuS~\cite{wang2021neus}. 3DGS~\cite{kerbl20233d} represents scenes as anisotropic Gaussians with learnable position, covariance, color, and opacity, achieving real-time differentiable rendering. Dynamic extensions include deformation-field approaches~\cite{wu20244d,yang2023deformable3dgaussianshighfidelity}, explicit motion models~\cite{wan2024superpoint}, and FEM mesh registration~\cite{shao2024gaussim}. We adopt 3DGS for its natural one-to-one correspondence with MPM particles.

Physics--rendering integration has advanced rapidly. PhysGaussian~\cite{xie2024physgaussian} warps Gaussian covariance via the deformation gradient to couple MPM with 3DGS; OmniPhysGS~\cite{lin2025omniphysgs} extended this to general constitutive models and Gaussian Splashing~\cite{feng2025gaussiansplashing} to fluid simulation. PAC-NeRF~\cite{li2023pacnerf} coupled NeRF with MPM for system identification, while PhysDreamer~\cite{zhang2024physdreamerphysicsbasedinteraction3d} and DreamPhysics~\cite{huang2024dreamphysics} estimated material properties from video/text priors to drive physics on 3DGS. Spring-Gaus~\cite{zhong2024reconstruction} attached Gaussians to spring-mass meshes for physics-based dynamic generation. However, these works all focus on forward simulation with learned material properties and do not support gradient-based inverse optimization toward a morphing target. Target-driven control of fluid simulations~\cite{mcnamara2004fluid,10.1145/1015706.1015743,10.1145/3016963} shares our goal but operates in Eulerian settings without the Lagrangian particle--Gaussian duality. Our work addresses the complementary inverse problem of steering physics toward a visual target.

\section{Method}
\label{sec:method}

\subsection{Overview}

Given a source shape (e.g., isosphere) and a target shape (e.g., the cow model), our goal is to produce a physically plausible morphing animation where the source continuously deforms into the target over a sequence of simulation frames. The source volume is discretized as a set of $N$ MPM particles, each carrying position $\mathbf{x}_i$, velocity $\mathbf{v}_i$, and deformation gradient $\mathbf{F}_i$. Each particle simultaneously serves as a 3D Gaussian primitive for differentiable rendering.

The optimization proceeds in frames, each consisting of a short MPM simulation ($T$ timesteps) followed by multi-view rendering and gradient extraction. The particle state is promoted across frames: positions and velocities at the end of frame $k$ become the initial conditions for frame $k\!+\!1$, creating a progressive morphing trajectory (Figure~\ref{fig:pipeline}).

Each frame performs three stages:
\begin{enumerate}
\item \textbf{Physics rollout.} An MPM forward simulation runs $T$ timesteps. A learnable control deformation field $\delta\mathbf{F}_c$ is optimized via the Adam optimizer~\cite{kingma2017adammethodstochasticoptimization} to minimize an end-layer mass loss that measures deviation from the target mass distribution on the Eulerian grid. If available, the render gradient $\partial \mathcal{L}_\text{render} / \partial \mathbf{F}$ from the previous frame is injected as a penalty on $\mathbf{F}$ (with the position component $\partial \mathcal{L} / \partial \mathbf{x}$ set to zero).

\item \textbf{Multi-view rendering and gradient extraction.} The final particle state $(\mathbf{x}_T, \mathbf{F}_T)$ is rendered from up to $V\!=\!8$ ring-placed cameras using the differentiable $\mathbf{F} \to \boldsymbol{\Sigma}$ mapping (Section~\ref{sec:duality}). Rendering losses (distance transform, soft IoU, and depth) are computed against target silhouettes obtained by rasterizing the target mesh. Backpropagation through the differentiable rendering chain yields $\partial \mathcal{L} / \partial \mathbf{F}$, which is stored for injection in the next frame.

\item \textbf{Chamfer plasticity injection.} Render injection provides lightweight $\mathbf{F}$-space guidance from the first frame, while Chamfer-guided plasticity activates only after a warmup of $k_0$ frames once coarse structural seeds have formed. This phased schedule ensures that rest-state migration refines rather than overrides early physical exploration (Section~\ref{sec:chamfer}).
\end{enumerate}

These three stages are complementary: render injection provides surface-level guidance, while plasticity persists achieved deformation across frames (detailed in Sections~\ref{sec:f_inject}--\ref{sec:chamfer}).

Algorithm~\ref{alg:pipeline} summarizes the complete pipeline.

\subsection{MPM--Gaussian Duality}
\label{sec:duality}

The central observation enabling our approach is a natural correspondence between the state variables of MPM simulation and the parameters of 3D Gaussian splatting. Each MPM particle $i$ carries position $\mathbf{x}_i \in \mathbb{R}^3$ and a deformation gradient $\mathbf{F}_i \in \mathbb{R}^{3 \times 3}$ that tracks how the local material neighborhood has been deformed from the rest configuration. In 3DGS, each Gaussian primitive is parameterized by a mean (position) $\boldsymbol{\mu}_i$ and a covariance matrix $\boldsymbol{\Sigma}_i \in \mathbb{R}^{3 \times 3}$ that determines its spatial extent and orientation.

We establish the duality $\boldsymbol{\mu}_i = \mathbf{x}_i$ (shared position) and derive $\boldsymbol{\Sigma}_i$ from $\mathbf{F}_i$ via the left Cauchy--Green deformation tensor~\cite{bonet1997nonlinear}. The Gaussian covariance is:
\begin{equation}
\boldsymbol{\Sigma}_i = \mathbf{F}_i \, \boldsymbol{\Sigma}_0 \, \mathbf{F}_i^\top
\label{eq:cov}
\end{equation}
where $\boldsymbol{\Sigma}_0 = \sigma_0^2 \mathbf{I}$ is the isotropic rest covariance. Since $\boldsymbol{\Sigma}_0$ is isotropic, this is equivalent to $\sigma_0^2 \mathbf{F}_i \mathbf{F}_i^\top$, capturing both stretch and rotation of the local material neighborhood. Following PhysGaussian~\cite{xie2024physgaussian}, this formulation maps the full deformation state to Gaussian shape without requiring explicit polar decomposition. The rest size $\sigma_0 = c \cdot d_\text{nn}$ is determined from the nearest-neighbor distance $d_\text{nn}$ with $c\!=\!0.7$.

The mapping is fully differentiable. Given a rendering loss $\mathcal{L}_\text{render}$, backpropagation yields:
\begin{equation}
\frac{\partial \mathcal{L}}{\partial \mathbf{F}_i} = \frac{\partial \mathcal{L}}{\partial \boldsymbol{\Sigma}_i} \frac{\partial \boldsymbol{\Sigma}_i}{\partial \mathbf{F}_i}
\label{eq:chain}
\end{equation}
This gradient modifies the deformation of particle $i$ to improve rendering quality without producing any direct gradient on position $\mathbf{x}_i$.

\begin{figure*}[t]
  \centering
  \includegraphics[width=0.8 \linewidth]{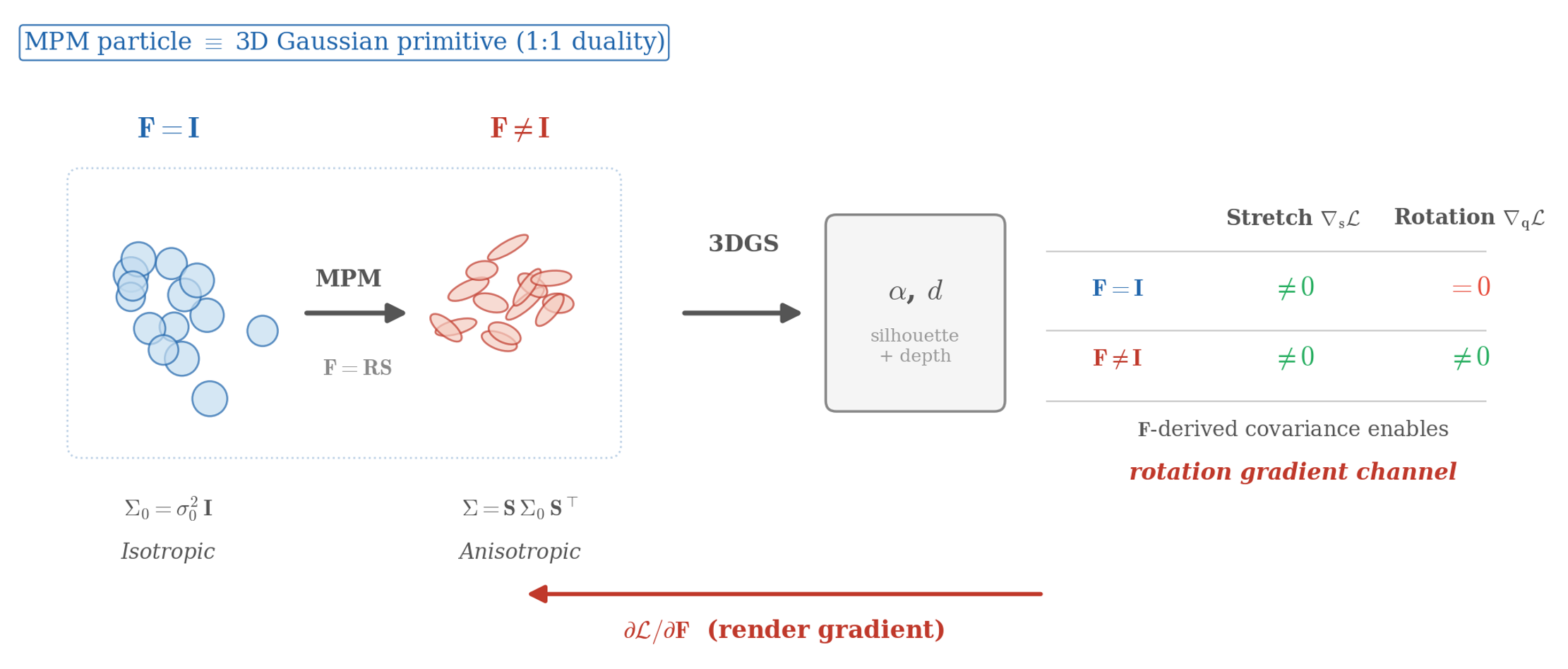}
  \caption{MPM--Gaussian duality. Each MPM particle's deformation gradient $\mathbf{F}_i$ determines the Gaussian covariance $\boldsymbol{\Sigma}_i = \mathbf{F}_i \boldsymbol{\Sigma}_0 \mathbf{F}_i^\top$. Render gradients $\partial \mathcal{L}/\partial \mathbf{F}$ flow back through this mapping, modifying the control deformation field without directly altering particle positions.}
  \label{fig:duality}
\end{figure*}

\subsection{Control-Space Render Injection}
\label{sec:f_inject}

The differentiable MPM framework supports injecting external gradients into the physics optimization loop as an additive penalty consisting of $(\mathbf{g}^\mathbf{F},\, \mathbf{g}^\mathbf{x})$, where $\mathbf{g}^\mathbf{F}_i \in \mathbb{R}^{3 \times 3}$ acts on the deformation gradient and $\mathbf{g}^\mathbf{x}_i \in \mathbb{R}^3$ acts on particle positions. We set:
\begin{equation}
(\mathbf{g}^\mathbf{F}, \; \mathbf{g}^\mathbf{x}) = \left( \gamma \, \frac{\partial \mathcal{L}_\text{render}}{\partial \mathbf{F}}, \; \mathbf{0} \right)
\label{eq:penalty}
\end{equation}
where $\gamma$ is the render-F gain (we use $\gamma\!=\!0.1$).

\subsubsection{Preferability of F-space injection} The position gradient $\partial \mathcal{L}_\text{render}/\partial \mathbf{x}$ lives in configuration space, whereas the MPM update is driven by stresses and thus by the deformation gradient $\mathbf{F}$. The two gradients have no natural magnitude correspondence, so mixing them as an additive penalty requires ad-hoc cross-space scaling; in our early experiments, unnormalized $\partial \mathcal{L}/\partial \mathbf{x}$ injection produced instability—including eigen-decomposition failures in the return mapping—at gains that $\partial \mathcal{L}/\partial \mathbf{F}$ absorbed without issue. Injecting the render signal through $\mathbf{F}$, by contrast, is dimensionally aligned with the Piola–Kirchhoff stress that the particle-to-grid transfer already integrates: the render signal composes directly with the physics step, with no cross-space magnitude matching required. A secondary benefit is coverage: $\partial \mathcal{L}/\partial \mathbf{x}$ is nonzero only on surface-adjacent particles (depth and silhouette losses are insensitive to interior positions), whereas $\partial \mathcal{L}/\partial \mathbf{F}$ propagates through every particle contributing to render-pixel covariances.

\subsubsection{Control-space injection mechanism} The injected $\partial \mathcal{L} / \partial \mathbf{F}$ is added to the physics gradient during the backward pass, influencing the Adam update of the control deformation field $\delta\mathbf{F}_c$. In the subsequent forward pass, the modified $\delta\mathbf{F}_c$ changes the stress distribution, which indirectly alters particle trajectories. This influence is mediated by the elastic energy landscape: the physics solver respects material properties, conservation laws, and stability constraints. Rather than applying an external force that directly opposes elastic restoring forces, we adjust the control signal so that the physics itself produces trajectories more consistent with the visual target.

The render gradient is computed at the end of frame $k$ and injected into the physics rollout of frame $k\!+\!1$. To prevent the injected signal from amplifying elastic oscillation, we apply velocity damping during the particle-to-grid (P2G) transfer ($\mathbf{p}_i \leftarrow m_i \mathbf{v}_i (1 - \Delta t \cdot d)$, $d\!=\!0.9$) and temporal deformation gradient smoothing ($\mathbf{F}_i^{t} \leftarrow (1-s)\mathbf{F}_i^{t} + s\mathbf{F}_i^{t-1}$, $s\!=\!0.955$, following Xu et al.~\cite{11088224}).

\subsection{Chamfer-Guided Plasticity}
\label{sec:chamfer}

Even with control-space render injection, physics-based morphing faces a fundamental stability challenge: the elastic energy $\Psi(\mathbf{F}_e)$ has a global minimum at $\mathbf{F}_e = \mathbf{I}$ (the undeformed rest configuration). As the optimizer pushes particles toward the target, accumulated elastic energy creates increasingly strong restoring forces, causing rebound, i.e., degradation of achieved shape quality as the system relaxes toward the source shape.

We address this through the multiplicative decomposition $\mathbf{F} = \mathbf{F}_e \mathbf{F}_p$~\cite{lee1969elastic}, where stress depends only on $\mathbf{F}_e = \mathbf{F} \mathbf{F}_p^{-1}$. Modifying $\mathbf{F}_p$ redefines the material's rest state without creating stress. If $\mathbf{F}_p$ is updated so that the current configuration becomes the new rest state, then $\mathbf{F}_e \approx \mathbf{I}$ and restoring forces point toward the target rather than back toward the source.

We drive the plasticity update using Chamfer nearest-neighbor correspondences~\cite{fan2017point}. While Chamfer distance can exhibit structural biases toward convex regions~\cite{song2026structuralfailurechamferdistance}, spatial diffusion of the displacement field mitigates this effect in our setting. At each frame after a warmup of $k_0$ frames, we: (1)~compute the nearest-neighbor displacement $\mathbf{d}_i = \mathbf{x}_i^\text{nn} - \mathbf{x}_i$ from each particle to the target surface via a K-D tree; (2)~smooth the displacement field using iterative KNN averaging ($k\!=\!64$, 3 iterations) to suppress noise; (3)~estimate the symmetric displacement Jacobian
\begin{equation}
\delta\mathbf{F}_p^{(i)} = \frac{1}{2}\left(\mathbf{J}_i + \mathbf{J}_i^\top\right), \quad \mathbf{J}_i = \frac{1}{|K_i|}\sum_{j \in K_i} \frac{(\mathbf{d}_j - \mathbf{d}_i)(\mathbf{x}_j - \mathbf{x}_i)^\top}{\|\mathbf{x}_j - \mathbf{x}_i\|^2}
\label{eq:dfp}
\end{equation}
and (4)~update multiplicatively: $\mathbf{F}_p \leftarrow (\mathbf{I} + \eta_i \, \delta\mathbf{F}_p) \, \mathbf{F}_p$, where $\eta_i$ is an adaptive rate scaled by local displacement magnitude (we use base $\eta\!=\!0.2$). To prevent unbounded accumulation, we apply a damping-toward-identity step $\mathbf{F}_p \leftarrow (1 - d_p)\mathbf{F}_p + d_p \mathbf{I}$ with $d_p\!=\!0.05$, followed by isochoric projection~\cite{simo1998computational} that bounds the determinant and anisotropy of $\mathbf{F}_p$.

The combined effect is progressive migration of the elastic equilibrium toward the target, converting physics from adversary to ally.

\subsection{Rendering Losses}
\label{sec:losses}

We render particles from 8 cameras and compute losses against target silhouettes. Our primary loss uses the signed distance transform~\cite{borgefors1986distance} of the target silhouette:
\begin{equation}
\mathcal{L}_\text{DT} = \frac{1}{|\Omega|} \sum_{p \in \Omega} \alpha_p \cdot \text{DT}(p)
\label{eq:dt}
\end{equation}
where $\alpha_p$ is the predicted alpha and $\text{DT}(p)$ is positive outside and negative inside the target boundary. Unlike binary cross-entropy (BCE), the DT gradient is proportional to boundary distance, providing a consistent directional ``pull'' that can be viewed as a projected Chamfer distance on the 2D image plane.

We supplement DT with soft IoU ($\mathcal{L}_\text{IoU} = 1 - \text{intersection}/\text{union}$) for global overlap and a depth loss for z-direction alignment. Adaptive gradient matching~\cite{yu2020gradient} normalizes auxiliary loss gradients to a target fraction of the DT gradient norm.

\subsection{Surface-Focused Rendering Mask}
\label{sec:surface}

A volumetric particle representation contains many interior particles that do not contribute to rendered images. In our implementation, we therefore use a surface-focused rendering mask to avoid spending rendering cost and gradient budget on visually irrelevant particles. We identify surface particles via density-based reconstruction on a $64^3$ voxel grid with Gaussian-filtered density, followed by marching cubes~\cite{lorensen1998marching} surface extraction and distance thresholding. Only particles within a threshold distance of the reconstructed surface participate in the 3DGS rendering pass and receive render gradients; the mask often filters out 60\%+ of particles. This practical masking choice focuses the gradient signal on visually relevant particles and reduces computation.

To further improve coverage, we augment the density-based mask with a Chamfer-based criterion: particles whose nearest-neighbor distance to the target surface falls below a threshold are progressively added (union, never shrink), ensuring that late-migrating particles still receive render gradients.

\begin{algorithm}[t]
\caption{PhysMorph-GS}
\label{alg:pipeline}
\begin{algorithmic}[1]
\REQUIRE Particles $\{(\mathbf{x}_i, \mathbf{v}_i, \mathbf{F}_i)\}_{i=1}^N$, target $\mathcal{T}$, cameras $\{C_v\}_{v=1}^V$
\STATE $\mathbf{F}_{p,i} \leftarrow \mathbf{I}$; \; $\mathbf{g}^{\mathbf{F}} \leftarrow \mathbf{0}$
\FOR{frame $k = 1, \ldots, K$}
    \STATE \textbf{// Stage A: Physics optimization}
    \STATE Update surface mask $\mathcal{S}$ via density recon.\ + Chamfer augmentation
    \FOR{GD iteration $j = 1, \ldots, J$}
        \FOR{substep $t = 1, \ldots, T$}
            \STATE $\mathbf{F}_{e,i} \leftarrow \mathbf{F}_i \, \mathbf{F}_{p,i}^{-1}$; \; $\mathbf{P}_i \leftarrow \text{PK1}(\mathbf{F}_{e,i})$ \COMMENT{MPM forward: PK1 = 1st Piola--Kirchhoff stress}
            \STATE P2G with drag: $m_i\mathbf{v}_i(1 - \Delta t\, d)$
            \STATE Grid solve; G2P update $\mathbf{v}_i$, $\mathbf{F}_i$
            \STATE $\mathbf{F}_i \leftarrow (1-s)\,\mathbf{F}_i + s\,\mathbf{F}_i^{t-1}$ \COMMENT{F smoothing}
        \ENDFOR
        \STATE Backward: $\nabla_{\delta\mathbf{F}_c} \mathcal{L}_\text{mass} + \gamma\,\mathbf{g}^{\mathbf{F}}$ \COMMENT{Render penalty in backward step}
        \STATE Adam step on $\delta\mathbf{F}_c$
    \ENDFOR
    \STATE \textbf{// Stage B: Multi-view rendering}
    \STATE $\boldsymbol{\Sigma}_i \leftarrow \mathbf{F}_i\,\sigma_0^2\mathbf{I}\,\mathbf{F}_i^\top$ for $i \in \mathcal{S}$
    \STATE Rasterize from $V$ views; compute $\mathcal{L}_\text{render}$
    \STATE $\mathbf{g}^{\mathbf{F}} \leftarrow \partial \mathcal{L}_\text{render} / \partial \mathbf{F}$ \COMMENT{Stored for frame $k\!+\!1$}
    \STATE \textbf{// Stage C: Chamfer plasticity} (if $k \geq k_0$)
    \STATE $\mathbf{d}_i \leftarrow \text{NN}_{\mathcal{T}}(\mathbf{x}_i) - \mathbf{x}_i$; smooth via KNN diffusion
    \STATE $\mathbf{F}_{p,i} \leftarrow (\mathbf{I} + \eta_i\,\delta\mathbf{F}_{p,i})\,\mathbf{F}_{p,i}$; damp toward $\mathbf{I}$; project
    \STATE Promote: initial state of $k\!+\!1$ $\leftarrow$ end state of $k$
\ENDFOR
\end{algorithmic}
\end{algorithm}

\section{Experiments}
\label{sec:experiments}

\subsection{Setup}

We conduct systematic ablations that isolate each proposed component, using the physics-only baseline of Xu et al.~\cite{11088224} as reference.

We use a differentiable MPM implementation with fixed corotated elasticity~\cite{2012-FixedCoratedElasticty}, chosen for its numerical robustness under large deformation and invertibility guarantees that prevent element inversion during aggressive morphing. Material parameters are fixed across all experiments (Young's modulus $E\!=\!1.4\!\times\!10^5$, Poisson ratio $\nu\!=\!0.2$, density $\rho\!=\!75$), as our focus is on the coupling strategy rather than material modeling. The simulation domain is $[-16, 16]^3$ with grid resolution $64^3$ ($\Delta x = 0.5$). Source shapes are sampled with shell-biased sampling (surface particle per cell (PPC)\,5, interior PPC\,1), yielding approximately 375K-748K particles per model. Time step $\Delta t = 1/240$\,s with $T\!=\!20$ timesteps per frame. We run 90 frames for main results and 60 frames for ablation studies. Velocity drag $d\!=\!0.9$ and deformation smoothing $s\!=\!0.955$ unless otherwise noted.

Eight cameras are placed in a ring configuration at $21^\circ$ elevation, rendering at $960 \times 540$. Gaussian size $\sigma_0$ is computed from nearest-neighbor spacing ($c\!=\!0.7$). We evaluate quantitatively on three isosphere-to-target benchmarks (Stanford Bunny, Cow, Duck), and additionally show cross-shape morphings (Duck$\to$Cow, Cow$\to$Bunny, Cow$\to$Duck) and letter-shape morphings (Isosphere$\to$A, Cow$\to$S, Duck$\to$C) qualitatively. Our primary metric is alpha MSE, measuring silhouette error against the rasterized target. We choose this metric because the method is supervised primarily through multi-view silhouette agreement, so alpha MSE directly reflects the quantity being optimized and compared across methods; we complement it with depth supervision during optimization and with qualitative render comparisons in the figures. Table~\ref{tab:params} summarizes all key parameters.

\begin{table}[t]
\centering
\small
\caption{Per-experiment parameters. All share: grid $64^3$, $\Delta t\!=\!1/240$\,s, $T\!=\!20$, $\gamma\!=\!0.1$, $\sigma_0$ scale $0.7$, Chamfer $\eta\!=\!0.2$, $k_0\!=\!20$, $E\!=\!1.4\!\times\!10^5$, $\nu\!=\!0.2$, $\rho\!=\!75$.}
\label{tab:params}
\begin{tabular}{lccc}
\hline
Experiment & Particles & $d$ & Frames \\
\hline
Iso $\to$ Bunny & 534K & 0.9 & 90 \\
Iso $\to$ Cow & 534K & 0.9 & 90 \\
Iso $\to$ Duck & 534K & 0.9 & 90 \\
Duck $\to$ Cow & 748K & 0.9 & 90 \\
Cow $\to$ Bunny & 545K & 0.9 & 90 \\
Cow $\to$ Duck & 545K & 0.75 & 90 \\
Iso $\to$ A & 534K & 0.9 & 90 \\
Cow $\to$ S & 375K & 0.9 & 90 \\
Duck $\to$ C & 454K & 0.9 & 90 \\
Ablation & 534K & 0.5 & 60 \\
\hline
\end{tabular}
\end{table}

\subsection{Main Results}

\begin{figure*}[t]
  \centering
  \includegraphics[width=0.85\linewidth]{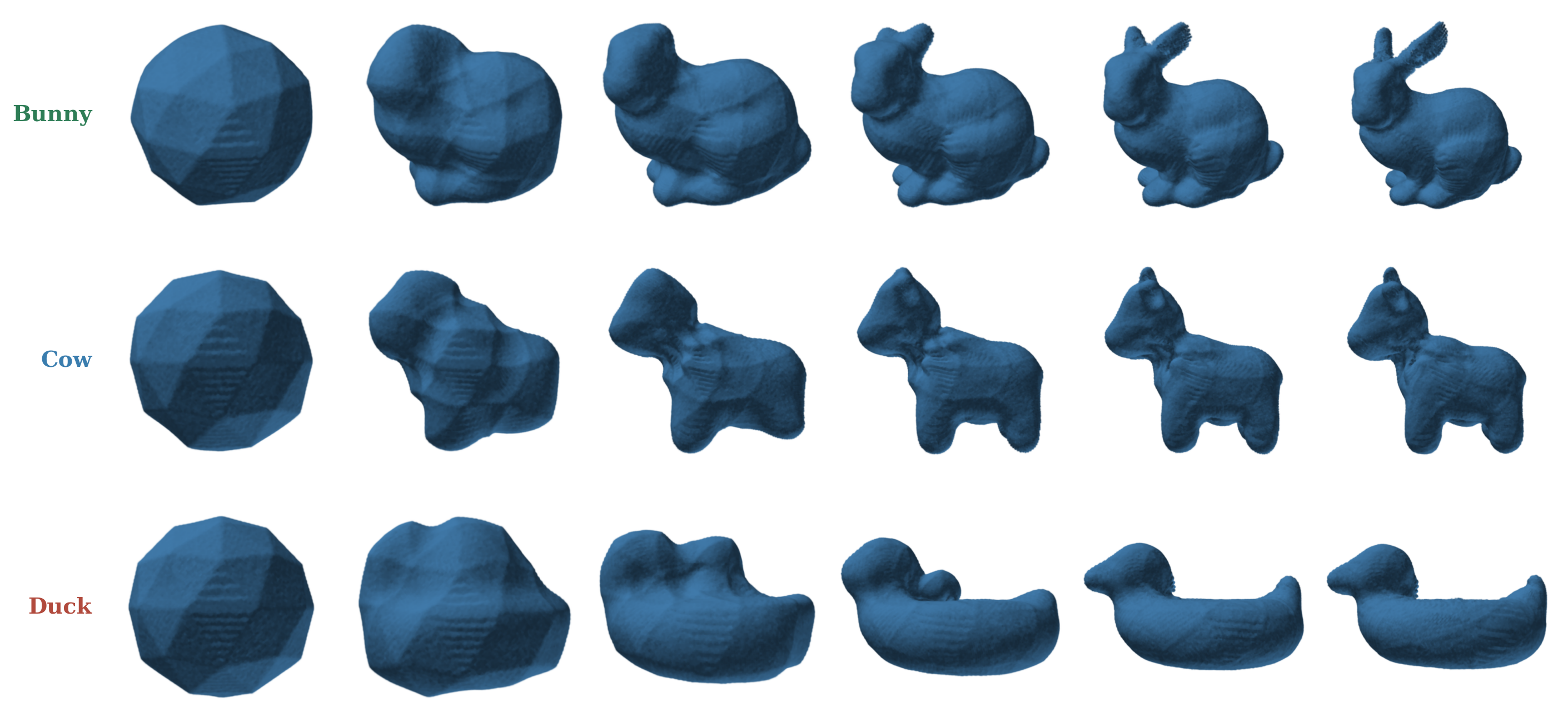}
  \caption{Isosphere-to-target morphing sequences. Each row shows progressive deformation from a common isosphere source to Bunny ($d\!=\!0.9$), Cow ($d\!=\!0.9$), and Duck ($d\!=\!0.9$). All particles rendered via 3DGS with Phong shading. Quantitative metrics in Table~\ref{tab:main}.}
  \label{fig:multi_target}
\end{figure*}

Table~\ref{tab:main} compares our method against the physics-only baseline at matched convergence. Rather than compare only the final frame, which would conflate shape quality with different convergence speeds and rebound behavior, we evaluate at the frame where physics-only has achieved $\geq$90\% of its total silhouette improvement. This gives a comparison at a similar stage of physics progress while still preserving the effect of render guidance on the trajectory. At this point, our method achieves 25.8\% lower silhouette error on Bunny, 10.8\% on Cow, and 49.9\% on Duck. The gain is largest on Duck, whose thinner protrusions benefit most from render guidance, while Cow shows the smallest but still consistent improvement because the physics-only baseline already converges relatively well on this more convex target. Figure~\ref{fig:multi_target} shows representative morphing sequences, and additional qualitative examples demonstrate that the same framework also extends to cross-shape and letter targets.

\begin{table}[t]
\centering
\caption{Morphing results ($\alpha$ MSE $\downarrow$). Comparison at matched physics convergence ($\geq$90\% of physics-only's total improvement, $d\!=\!0.9$), where our method consistently achieves lower silhouette error across all targets.}
\label{tab:main}
\begin{tabular}{lccc}
\hline
& Bunny & Cow & Duck \\
\hline
Physics-only & 0.0799 & 0.0777 & 0.0689 \\
Ours (same frame) & \textbf{0.0593} & \textbf{0.0694} & \textbf{0.0345} \\
Improvement & 25.8\% & 10.8\% & 49.9\% \\
\hline
\end{tabular}
\end{table}

\subsection{Cross-Shape Morphing}
\label{sec:cross_shape}

\begin{figure*}[t]
  \centering
  \includegraphics[width=0.85\linewidth]{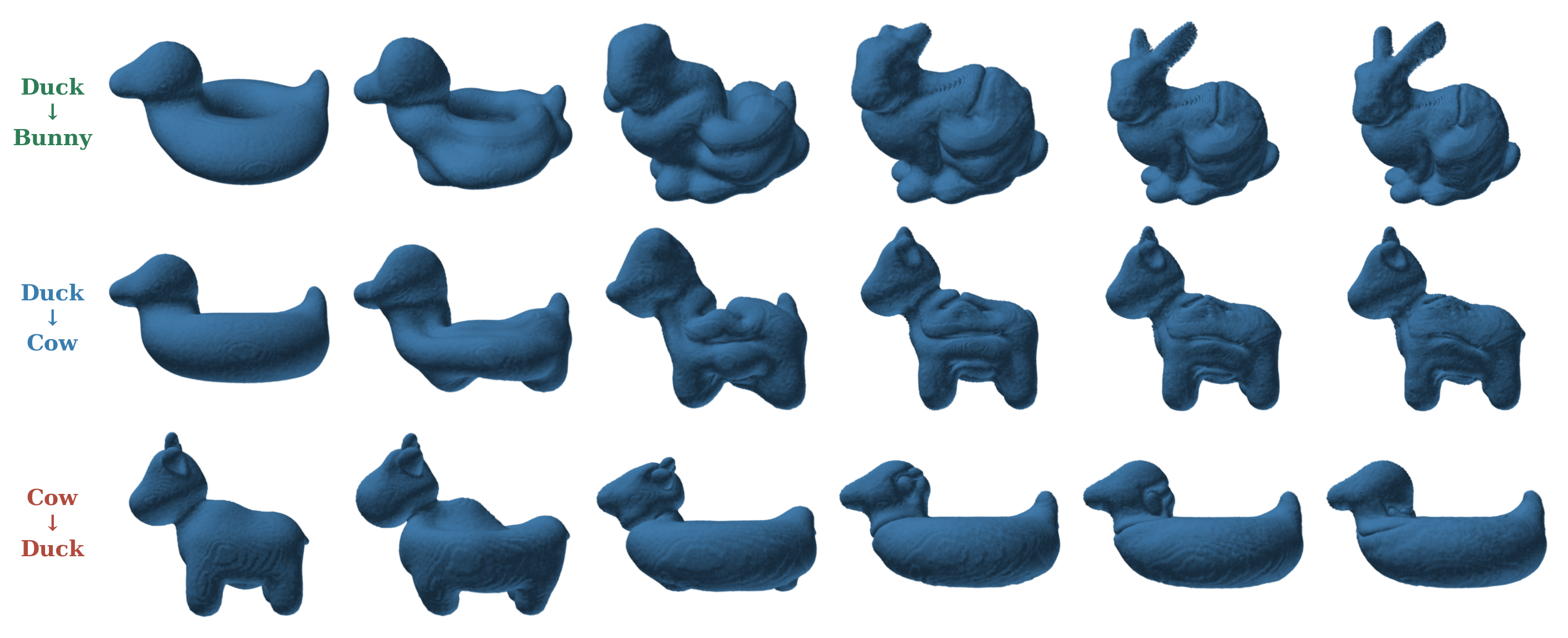}
  \caption{Cross-shape morphing sequences. Rows: Duck$\to$Bunny, Duck$\to$Cow, Cow$\to$Duck. Each row samples six frames from $t\!=\!0$ (source) through the final morphed target. The same pipeline (F-space render injection, phased Chamfer plasticity, $d\!=\!0.9$) handles bidirectional transitions between animal shapes with non-convex features (limbs, ears, beaks) without per-pair tuning.}
  \label{fig:cross_shape}
\end{figure*}

Beyond the isosphere-to-target setting used for quantitative comparison, the same framework also produces plausible morphings between non-trivial source and target pairs, including animal-to-animal transitions. Figure~\ref{fig:cross_shape} shows three such cases: Duck$\to$Bunny, Duck$\to$Cow, and Cow$\to$Duck. These morphings require the control deformation field and Chamfer-guided plasticity to simultaneously contract, extend, and reshape non-convex features (e.g., ears, limbs, beaks) across substantially different topologies. No additional tuning is required: the same pipeline parameters used for isosphere-to-target runs yield coherent cross-shape trajectories, suggesting that the F-space coupling and phased plasticity generalize beyond the canonical convex-to-target setting.

\subsubsection{Source invariance}
\label{sec:source_invariance}
In contrast to the previous cross-shape runs, which vary the target, here we fix the target as the Bunny model and vary the source. A distinguishing property of physics-driven rest-state migration is that the final morph is dominated by the target rather than the source. We evaluate this by morphing three topologically different sources---Isosphere (convex), Cow (quadruped), and Duck (limbless animal)---to the same Bunny target under identical pipeline parameters. The best-frame $\alpha$~MSE values are $0.0546$, $0.0460$, and $0.0558$ respectively (mean $0.0521$, $\sigma\!=\!0.0054$, $\sigma/\mu \approx 10.3\%$), indicating convergence to a target-dominated attractor with modest residual source-dependence. Visually, all three final states are recognizable as the target Bunny (see supplementary video), with differences concentrated in fine-scale features (e.g., ear tip sharpness, torso fullness, and residual surface noise from particles transitioning to the Gaussian-rendered surface late in the trajectory) that reflect the distinct deformation histories of each source rather than global structure. This property follows from the plasticity formulation: as $\mathbf{F}_p$ migrates to encode the target geometry in the rest configuration, the influence of the initial source state is progressively absorbed. Geometric interpolation methods that blend between registered source and target representations do not share this property by construction, as their trajectories are tied to a specific source--target correspondence.

\subsection{Ablation Studies}
\label{sec:ablation}

Our ablations focus on the two central questions raised by the method design: whether render guidance should be injected in control space rather than position space, and whether damping plus phased plasticity are needed to turn short-term deformation into stable progress.

\subsubsection{Coupled injection and damping}
Control-space render injection and velocity damping address complementary aspects of the morphing problem. Injection provides a directional signal that steers deformation toward the visual target, while damping ensures that this extra guidance is absorbed stably into the elastic simulation. Table~\ref{tab:ablation} shows that injection alone (at default $d\!=\!0.5$) improves final alpha MSE by 9.7\% on Bunny over the physics-only baseline, while also improving physics loss reduction (93.8\%$\to$95.9\%). However, the injected render signal can amplify elastic oscillation when the control field overshoots.

\begin{figure*}[t]
  \centering
  \includegraphics[width=0.85\linewidth]{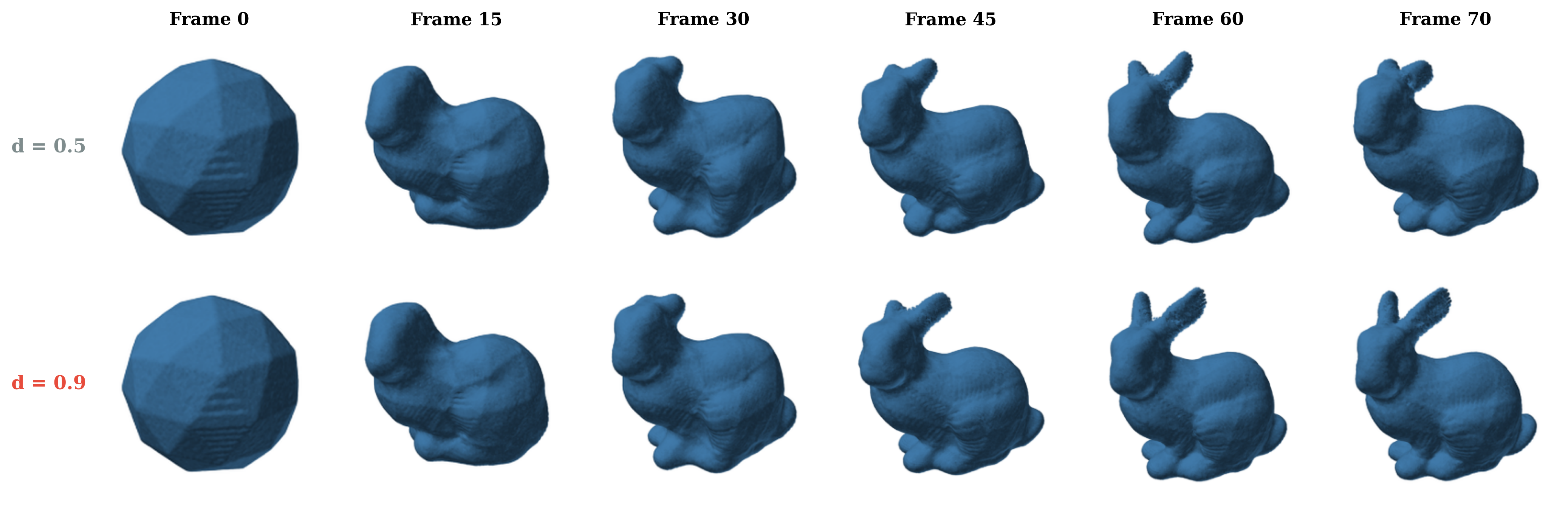}
  \caption{Effect of velocity damping on Isosphere$\to$Bunny morphing. Top row: default drag ($d\!=\!0.5$), bottom row: high drag ($d\!=\!0.9$, ours). With $d\!=\!0.5$, the shape oscillates and fails to converge cleanly. With $d\!=\!0.9$, the morphing proceeds smoothly, yielding 10.3\% lower silhouette error compared to physics-only at the same damping.}
  \label{fig:drag_ablation}
\end{figure*}

Increasing the drag coefficient to $d\!=\!0.9$ suppresses this oscillation (Figure~\ref{fig:drag_ablation}): the alpha MSE curve descends more monotonically rather than oscillating around the minimum. Crucially, comparing against a physics-only baseline at the same $d\!=\!0.9$ isolates the effect of render guidance from damping: in this damping-matched ablation on Bunny (best-achieved alpha MSE over 90 frames, Table~\ref{tab:ablation}), our full pipeline achieves 10.3\% lower error (0.0546 vs.\ 0.0609), confirming that the improvement over physics-only is not merely due to stronger damping. Note that this 10.3\% figure is a \emph{damping-matched} ablation comparing best-frame alpha MSE; the larger 25.8--49.9\% gains reported in Table~\ref{tab:main} are at matched physics convergence ($\geq\!90\%$), which is the setting we use for the main evaluation.

We do not treat direct position-space injection as a competitive optimization baseline. In practice it behaves as a negative control: the injected signal is sparse and noisy due to visibility, depth ambiguity, and the lack of supervision on interior particles, while also directly conflicting with elastic restoring forces on the same state variable. For this reason, our quantitative ablations focus on the stable control-space formulation and use the observed oscillatory behavior of direct image-space coupling as the motivating failure mode.

\begin{table}[t]
\centering
\caption{Ablation: silhouette error ($\alpha$ MSE $\downarrow$) and relative reduction of the mass-matching physics loss ($\Delta E$ \% $\uparrow$). Component analysis is reported on Bunny (PPC6, $32^3$ grid, 60 frames). Drag comparison is reported on Bunny (PPC5, $64^3$ grid, 90 frames).}
\label{tab:ablation}
\begin{tabular}{lcc}
\hline
\multicolumn{3}{c}{\textit{Component analysis (Bunny, 60 frames)}} \\
Method & $\alpha$ MSE & $\Delta E$ (\%) \\
\hline
Physics-only & 0.1152 & 93.8 \\
+ F-injection & 0.1043 & 95.8 \\
+ Chamfer plast.\ (ours) & \textbf{0.1040} & \textbf{95.9} \\
\hline
\multicolumn{3}{c}{\textit{Velocity damping (Bunny, 90 frames)}} \\
Method & $\alpha$ MSE (best) & $\alpha$ MSE (final) \\
\hline
Physics-only ($d\!=\!0.9$) & 0.0609 & 0.0741 \\
Full ($d\!=\!0.5$) & 0.0735 & 0.0786 \\
Full ($d\!=\!0.9$, ours) & \textbf{0.0546} & \textbf{0.0637} \\
\hline
\end{tabular}
\end{table}

\subsubsection{Trajectory divergence from F-injection} 
\begin{figure}[t]
  \centering
  \includegraphics[width=0.85\linewidth]{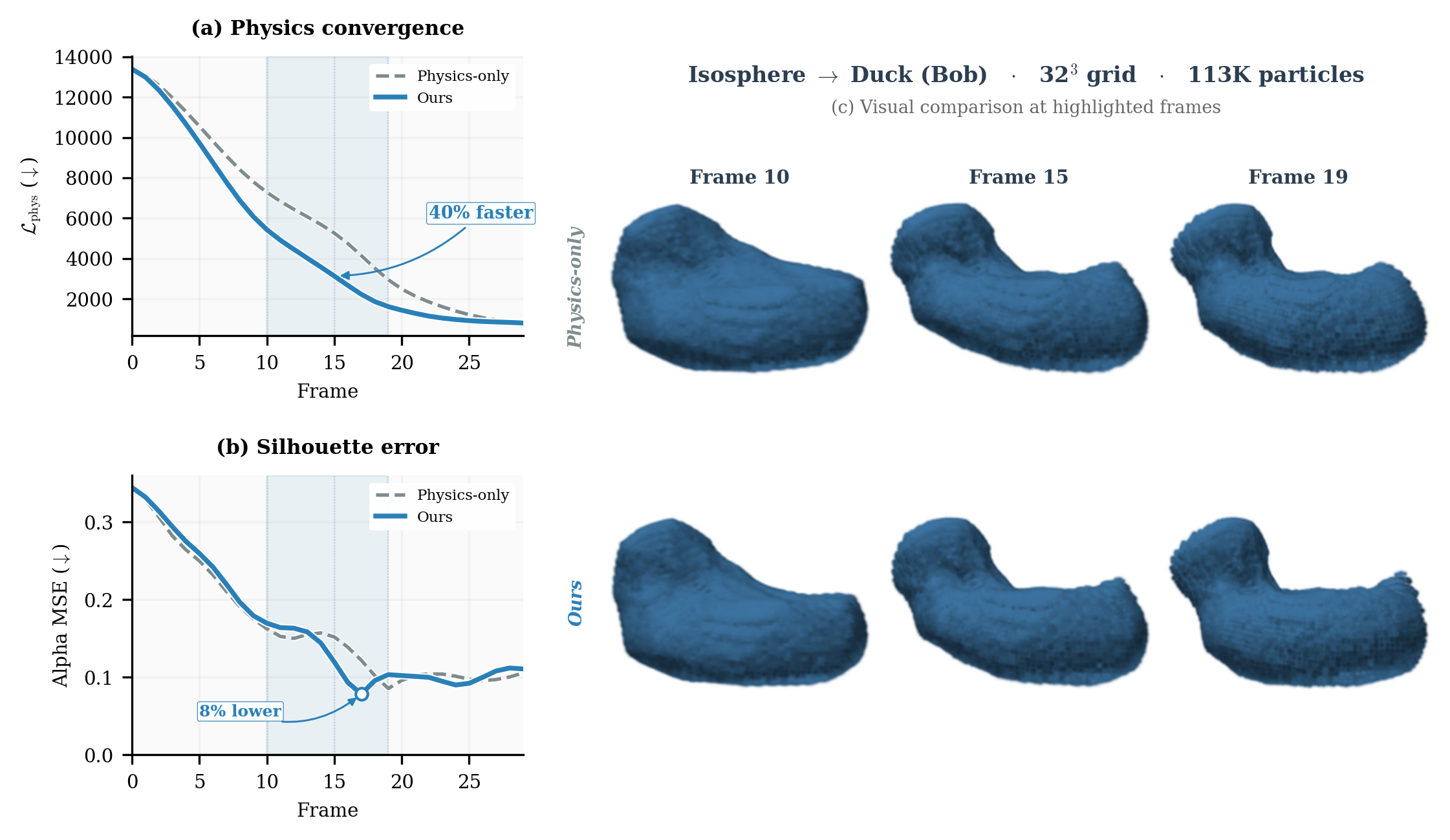}
  \caption{Trajectory divergence from F-injection on Isosphere$\to$Duck. \textbf{(a)}~Physics convergence: our method reaches comparable loss 40\% faster. \textbf{(b)}~Silhouette error: 8\% lower with render guidance. \textbf{(c)}~Visual comparison at highlighted frames shows qualitatively different intermediate shapes.}
  \label{fig:trajectory}
\end{figure}

Figure~\ref{fig:trajectory} illustrates how control-space render injection alters the morphing trajectory on Isosphere$\to$Duck. The physics convergence curve shows that our method reaches comparable physics loss 40\% faster, while the silhouette error diverges visibly from frame~10 onward. The accompanying renders at frames 10, 15, and 19 show that F-injection does not merely accelerate the baseline trajectory; it changes the intermediate shapes qualitatively, producing target-consistent curvature earlier instead of remaining in a flat blob-like configuration.

\subsubsection{Chamfer plasticity}
Without plasticity, elastic energy accumulates and produces rebound after the best shape quality is reached.
With Chamfer plasticity ($\eta\!=\!0.2$, $k_0\!=\!20$), the update of $\mathbf{F}_p$ absorbs deformation into the rest configuration, converting elastic restoring forces from adversary to ally.
As a result, rebound is reduced to less than $8\%$, while the physics loss trajectory remains essentially unchanged because modifying $\mathbf{F}_p$ does not directly introduce additional stress.

\begin{figure}[t]
  \centering
  \includegraphics[width=\linewidth]{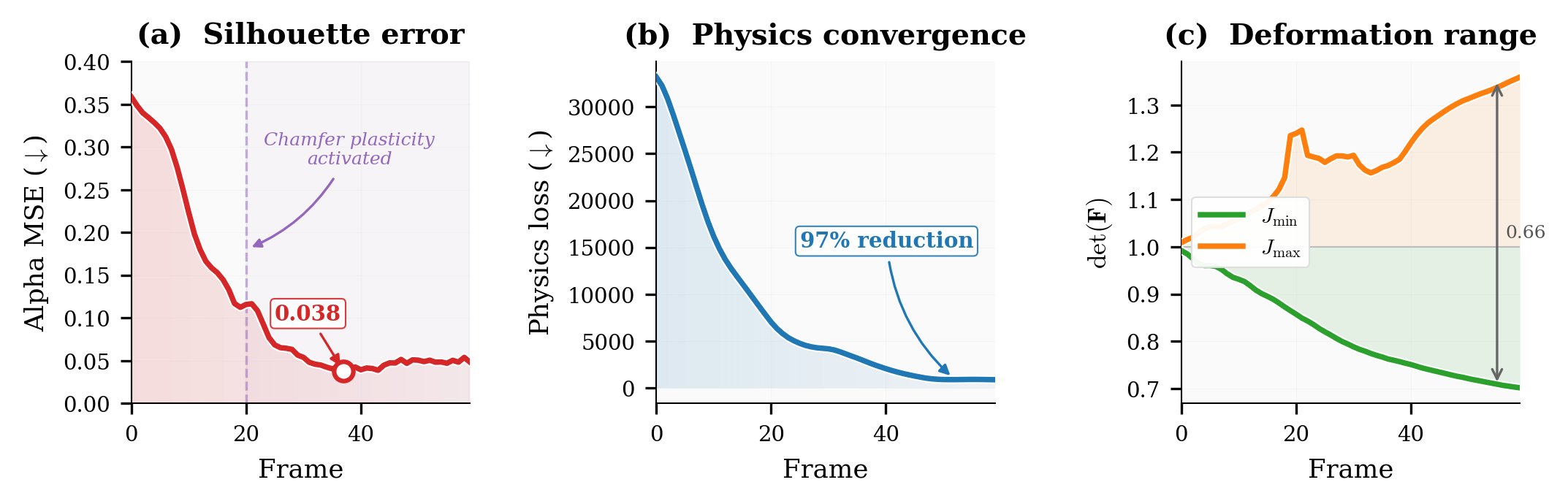}
  \caption{Effect of Chamfer-guided plasticity.
  (a) Silhouette error decreases further after plasticity is activated.
  (b) Physics loss continues to converge smoothly, indicating that the underlying simulation remains stable.
  (c) The deformation range stays bounded throughout the sequence, showing that the update to $\mathbf{F}_p$ does not introduce excessive distortion.}
  \label{fig:chamfer_plasticity}
\end{figure}

Direct Chamfer optimization can produce many-to-one correspondences that lead to particle clustering and loss of spatial coverage~\cite{song2026structuralfailurechamferdistance}, so we use Chamfer only to update $\mathbf{F}_p$ rather than as a direct objective.
As shown in Figure~\ref{fig:chamfer_plasticity}, this improves silhouette alignment while keeping the physics loss smooth, and the deformation bounded.

\begin{figure}[t]
  \centering
  \includegraphics[width=\linewidth]{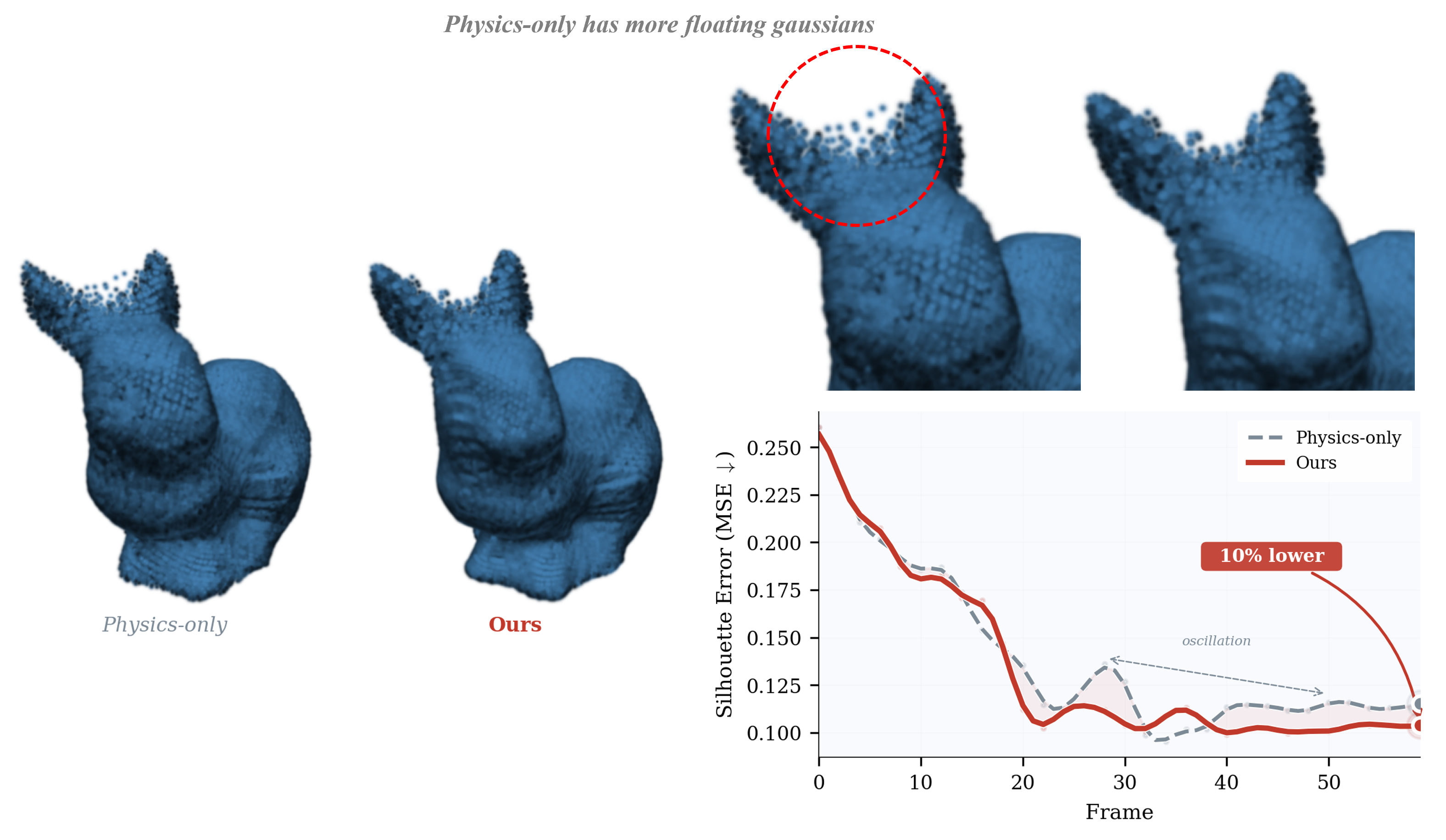}
  \caption{Component ablation on Bunny ($d\!=\!0.5$, 60 frames). Left: physics-only result with floating particles. Right: adding F-injection and Chamfer plasticity produces cleaner geometry. Graphs show silhouette error (9.7\% lower) and physics loss convergence.}
  \label{fig:ablation_render}
\end{figure}

\begin{figure}[t]
  \centering
  \includegraphics[width=\linewidth]{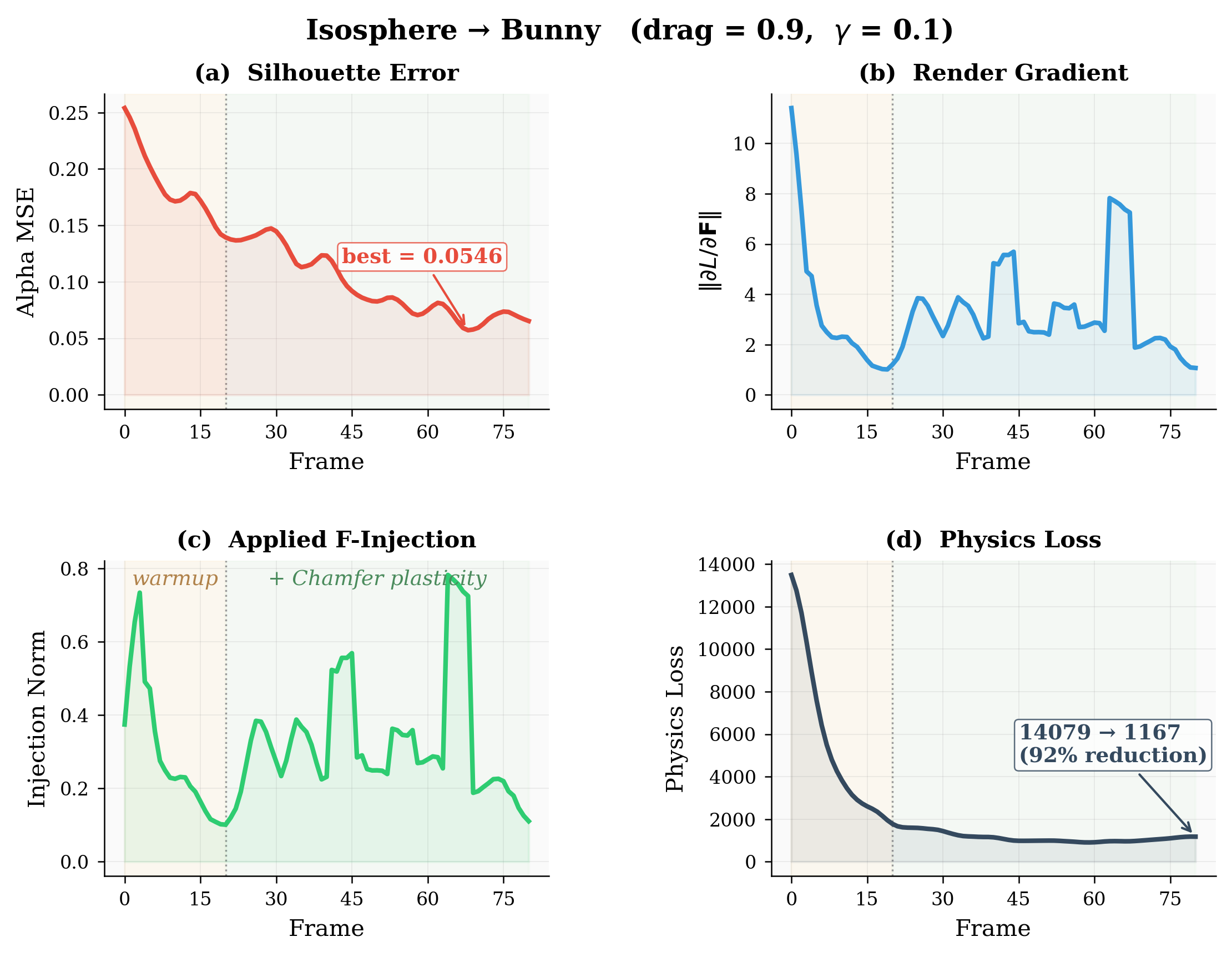}
  \caption{Optimization traces for Isosphere$\to$Bunny ($d\!=\!0.9$, $\gamma\!=\!0.1$): \textbf{(a)} alpha MSE reaches $0.0546$; \textbf{(b)} render gradient norm $\|\partial \mathcal{L}/\partial \mathbf{F}\|$ remains active past the initial transient; \textbf{(c)} applied F-injection, with dashed line marking Chamfer plasticity onset at $k_0\!=\!20$; \textbf{(d)} physics loss drops $14079\!\to\!1167$ (92\% reduction).}
  \label{fig:grad_tracking}
\end{figure}

\subsubsection{Multi-view coverage} 
\begin{figure}[t]
  \centering
  \includegraphics[width=\linewidth]{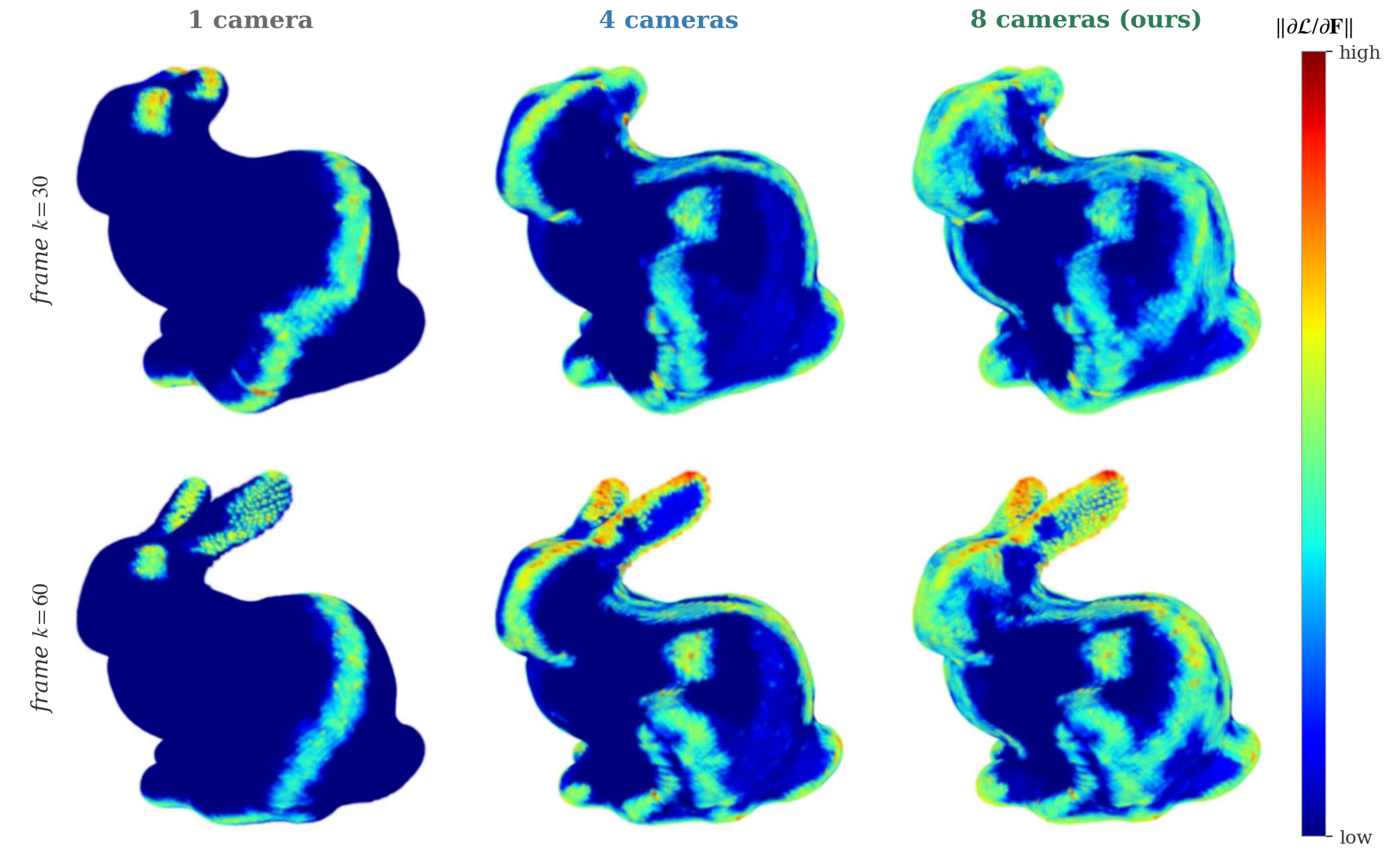}
  \caption{Per-particle render gradient magnitude ($\|\partial \mathcal{L}/\partial \mathbf{F}\|$, log scale) on the same Bunny state (frame 60) with 1, 4, and 8 cameras. Warm colors indicate high gradient; cool colors indicate low gradient. More cameras provide broader surface coverage for the render signal.}
  \label{fig:grad_coverage}
\end{figure}

Figure~\ref{fig:grad_coverage} visualizes the per-particle render gradient magnitude $\|\partial \mathcal{L}/\partial \mathbf{F}_i\|$ under different camera configurations. A single camera provides gradient only on the visible silhouette boundary, leaving the opposite side unguided. Four cameras improve lateral coverage but leave gaps between views. Eight cameras in a ring configuration provide full surface coverage, with gradient naturally concentrating on fine structures (ears, feet) where the predicted and target silhouettes disagree most.

\subsection{Timing}

Despite coupling differentiable MPM over 534K particles with multi-view Gaussian rendering and Chamfer plasticity, each frame completes in approximately 4 minutes on a single RTX GPU: physics rollout ($\sim$3 min), multi-view rendering and backward pass ($\sim$40 s), and Chamfer plasticity ($\sim$20 s).

\section{Discussion and Limitations}
\label{sec:discussion}


Although our algorithm performs well on morphing tasks between various source and target shapes, we note several limitations of the present approach.
First, particles cannot be created or destroyed during simulation, limiting resolution at thin protrusions where the source has insufficient density; adaptive refinement~\cite{adams2007adaptively,ando2012highly} could address this. Second, the surface mask is progressively augmented but particles migrating to the surface late may still miss render gradients.
Third, concave targets (e.g., letter shapes with interior holes) remain challenging because elastic forces resist the topology change required to evacuate interior regions, and multi-view rendering provides weak gradients in narrow concavities. Fourth, topology-closing regions can exhibit surface artifacts in the raw particle view, as contact mediated only through deformation and plasticity does not produce a clean watertight merge without downstream surfacing. Fifth, particle compression produces surface wrinkling, mitigated via Laplacian smoothing~\cite{taubin1995signal} for visualization (metrics remain on unsmoothed results). Sixth, Chamfer correspondences can exhibit structural bias toward convex regions~\cite{song2026structuralfailurechamferdistance,wu2021density}, although spatial diffusion partially mitigates this effect in practice. Finally, volume is not strictly conserved during morphing: while the isochoric projection on $\mathbf{F}_p$ bounds plastic volume change, elastic compression ($J < 1$) can still reduce apparent volume in high-curvature regions. Unlike purely geometric morphing methods that have no mechanism to track volume at all, our formulation exposes volume behavior as a measurable quantity that can be further constrained in future iterations. Incorporating a global volume penalty is a natural direction for future work.

\section{Conclusion}
\label{sec:conclusion}

We presented PhysMorph-GS, a framework for render-guided physics-based volumetric morphing with differentiable 3D Gaussian splatting. The central insight is to inject visual supervision through the deformation gradient $\mathbf{F}$ rather than particle positions, which makes render guidance compatible with elastic simulation instead of directly competing with it. Phased Chamfer-guided plasticity then migrates the rest configuration so that the achieved deformation can persist across frames. At matched physics convergence ($\geq$90\%), the resulting system reduces silhouette error by 25.8\%, 10.8\%, and 49.9\% on Bunny, Cow, and Duck, respectively, with the largest gains on targets containing finer structures. Beyond these benchmark results, the same formulation also produces plausible cross-shape and letter-shape morphings, and drives topologically distinct sources toward a shared target-determined attractor---a property that follows from plasticity-driven rest-state migration rather than source-to-target interpolation. Future work includes adaptive particle redistribution for thin structures, topology-aware Chamfer correspondences, stronger volume control, and extension to multi-material morphing.

\section{Acknowledgments}
\label{sec:acknowledgement}

Claude Opus 4.6 was used in the preparation of this work for coding and editing tasks.

\bibliographystyle{eg-alpha-doi}
\bibliography{egbibsample}
\end{document}